\documentclass[aps,prl,twocolumn,groupedaddress]{revtex4}
\usepackage{epsfig}
\usepackage{amsfonts}
\newcommand{\be}{\begin{equation}}
\newcommand{\ee}{\end{equation}}
\newcommand{\ba}{\begin{eqnarray}}
\newcommand{\ea}{\end{eqnarray}}

\begin{document}

\title{On three papers by Jurgens \& Crutchfield, and on the basic structure of ``computational mechanics"}

\author{Peter Grassberger}

\affiliation{JSC, FZ J\"ulich, D-52425 J\"ulich, Germany}

\date{\today}

\begin{abstract} In a recent paper, Jurgens and Crutchfield [Phys. Rev. E {\bf 104}, 064107 (2021), called 
``paper III" in the following]  computed what they called the ``ambiguity rate" of hidden Markov processes, 
a concept supposedly introduced by Claude Shannon. This calculation was based on a ``mixed state" formalism 
introduced by them in J. Stat. Phys. {\bf 183}, 32 (2021) (``paper I"), and developed further in 
Chaos, {\bf 31}, 083114 (2021) (``paper II"). We point out that (i) ambiguity rates were {\it not} introduced by 
Shannon; (ii) their computations in paper III are wrong, because of an error made already in papers I and II;
(iii) due to this error (a confusion between open sets and their closures), also many of the ``statistical complexity 
dimensions" computed in II are wrong; (iv) the ``mixed state" formalism of I is just the well 
known `forward algorithm' for hidden Markov models; (v) the `causal states' in `$\epsilon$-machines' 
correspond in general to {\it finite} (as opposed to infinite, as often claimed) histories; and (vi) 
`$\epsilon$-machines' are always countable, in contrast to frequent claims in the literature. In addition, 
we propose an alternative complexity measure for models where the 
forecasting complexity is infinite, and we point out that our results apply also beyond hidden Markov models.
\end{abstract}
\maketitle

Hidden Markov models (HMMs) \cite{Bengio,Ephraim,Zucchini} are a standard tool in data analysis, with
a vast number of applications covering among others finance, medicine, chemistry, bioinformatics, speech
analysis, meteorology, and physics. It is repeatedly claimed in the literature that no good algorithms
exit for estimating their (Shannon) entropy \cite{Jacquet,Ordentlich}, but this is not true: both the 
``forward" and the Baum-Welch algorithms \cite{Zucchini} do just this. It is true that in their treatments 
entropy is often not mentioned, but likelihoods are, and entropy is just the average negative log-likelihood.

In \cite{Jurgens1} (paper I), a supposedly new algorithm based on ``mixed states" was proposed. But this 
algorithm is exactly the forward algorithm, and the ``mixed states" become precisely the causal states
introduced in \cite{Grass86}, when their set is minimized. This minimization is, in general, anyhow necessary
for an efficient application. Indeed, the forward algorithm was independently (re-)discovered for a special
class of HMMs in \cite{Zambella}, where it was used to present precise entropy estimates for a number
of non-trivial models, and where it was also used to compute forecasting complexities (FCs \cite{Grass86}; later 
called `stochastic complexities' in \cite{Young}). The fact that the algorithm proposed in \cite{Zambella}
is much more general was stressed in \cite{arXiv}, where also a number of other points were clarified.

More precisely, we consider a finite stationary ergodic irreducible first order Markov process $M$. Here, 
``finite" means that it has a finite number $N$ of states, in the following called $s_t$ with values 
numbered simply by $``i",\; i = 1\ldots N$. The restriction to first order is trivial, as any higher order 
Markov chain can be re-written as a first order process by re-labelling the states. Irreducibility is
essential since it implies (by the Frobenius-Perron-theorem) that there is a unique stationary
state. The probability to be in state $i$ at (discrete) time $t$ is denoted as $P_{i,t}$. The
stationary state is characterized by the probabilities $P_{i,\infty}$. The transition matrix is
$T_{ji} = {\rm prob}\{s_{t+1}=j|s_t=i\}$. The time evolution of the probabilities is
\be
   P_{j,t+1} = \sum_i T_{ji} P_{i,t}.
\ee
In a more compact vector notation, we denote the state at time $t$ by a vector ${\bf P}_t\in \mathbb{R}^N$,
so that the time evolution can be written as ${\bf P}_{t+1} = {\bf T} {\bf P}_t$.

This Markov process is ``hidden", i.e. its states cannot be observed. What is observed instead
is the chain of discrete symbols $x_t$ which are emitted one-by-one during each transition.
These emitted symbols are chosen from an alphabet $A$ (with cardinality $|A|$), and
\be
   T^{(x)}_{ji} \equiv ({\bf T}^{(x)})_{ji} = {\rm prob}\{s_{t+1}=j, x_t = x|s_t=i\}.
\ee
For consistency, we have
\be
   {\bf T} = \sum_{x\in A} {\bf T}^{(x)}.
\ee
Notice that we consider thus ``edge-emitting" HMMs. The more widely discussed ``node-emitting" HMMs
\cite{Zucchini}
are obtained simply by assuming that the emitted symbol does depend only on state $i$ and not on $j$,
so that $T^{(x)}_{ji} = T_{ji} q(x|i)$.

We now assume that the process has been going on for a long time, but no emitted symbols have been 
observed for any time $t\leq 0$, so that ${\bf P}_0 = {\bf P}_\infty$. Thus the observer/forecaster
has at time $t=1$ no clue about the internal state of $M$, and cannot make any non-trivial
forecast beyond that based on stationarity. This changes, however, when observations are made
at times $t = 1,2,\ldots$. From these, the observer can make inferences about the internal states,
which will then lead to improved forecasts. Comparing these forecasts with the actually
emitted symbols will lead to improved inferences, and so on. In general, this {\it `observation 
process'} is {\it not} stationary (even if the HMM is). It will in the limit $t\to\infty$
lead to a situation where the non-predicted amount of information per symbol is a.s. the entropy.

The optimally inferred states of the hidden dynamics based on the observation process were called 
`mixed states' in \cite{Jurgens1,Jurgens2,Jurgens3} (papers I-III), and will be denoted here as 
$\hat{\bf P}_t(x_1x_2\ldots x_t)$. Based on them, the optimal forecast of the observed process is
$\hat{p}_{t+1}(x|x_1x_2\ldots x_t) = \hat{p}_{t+1}(x_1x_2\ldots x_t,x)/\hat{p}_t(x_1x_2\ldots x_t)$
with
\ba
   \hat{p}_{t+1}(x_1x_2\ldots x_t,x) & = &\sum_{i,j} T^{(x)}_{ji}\hat{P}_{i,t}(x_1x_2\ldots x_t) \\
               & = & {\bf 1} \cdot {\bf T}^{(x)} \hat{\bf P}_t(x_1x_2\ldots x_t)   \nonumber
               \label{forecast}
\ea
with ${\bf 1} = (1\ldots1)^T$. On the other hand, an unbiased hidden state inferred from a new
observation $x_{t+1}$ is
\be
    \hat{\bf P}_{t+1}(x_1x_2\ldots x_{t+1}) = c \cdot {\bf T}^{(x_{t+1})} \hat{\bf P}_t(x_1x_2\ldots x_t),
\ee
where the constant $c$ is determined by the normalization
${\bf 1} \cdot \hat{\bf P}_{t+1}(x_1x_2\ldots x_{t+1})=1$, which gives finally
\be
    \hat{\bf P}_{t+1}(x_1x_2\ldots x_{t+1}) =
    \frac{{\bf T}^{(x_{t+1})} \hat{\bf P}_t(x_1x_2\ldots x_t)}{\hat{p}_{t+1}(x_{t+1}|x_1x_2\ldots x_t)}.
    \label{conj}
\ee
The complete observation / forecasting process is thus a cyclic chain between inferences,
forecasts, and updates due to observations.

Equations (4) and (6) are essentially the same as Eq.(3.4) in \cite{Zambella}, and are today well
known as `forward algorithm' \cite{Zucchini}. They hold also for `observable operator models' (OOMs)
\cite{Jaeger,Spanczer} with discrete time and discrete values (in which the positivity requirements 
made for HMMs are dropped), and even for `matrix product states' in statistical physics 
\cite{Derrida,Klumper}).

While the Baum-Welch algorithm is often considered as superior to the forward algorithm, it is 
shown in \cite{Zucchini} that both have their advantages. Indeed, the entropy estimates of simple
but non-trivial models made in \cite{Zambella} were of unprecedented accuracy, and no brute-force 
estimates would have been possible at that time with the same accuracy. Using the estimator of 
\cite{arXiv2003}, we now checked that all entropies quoted in Table 1 of \cite{Zambella}
agreed perfectly with brute-force estimations, with one single exception: For rules 25 and 61,
the correct value of the entropy is 0.88955 bits, while 0.88947 bits was quoted in \cite{Zambella}.

As shown in \cite{Zambella} (see also \cite{arXiv}), the set of `mixed states' becomes precisely the 
set of causal states introduced in \cite{Grass86} when minimized. Minimization means that two 
vectors $\hat{\bf P}_t(x_1\ldots x_t)$ and $\hat{\bf P}_t(x'_1\ldots x'_t)$ are identified if they 
lead, via Eq.(4), to the same $\hat{p}_{t+k}(x) \; \forall k\geq 1$. No such minimization was done in 
I, which is why no FCs were estimated there and no precise estimates of entropy. 
It was also shown in \cite{Zambella} that causal states do in general {\it not} correspond to elements 
of a partitioning of the space of complete histories (in contrast to what was claimed e.g. in 
\cite{Crutch_physica,Shalizi,Feldman,Loomis}), at least when transient parts of the ``$\epsilon$-machine" are not cut 
off. Cutting them off (as proposed in most recent papers by Crutchfield {\it et al.}) is not a viable 
alternative, as in some cases (called `{\it totally recurrent graphs'} in \cite{arXiv}) none of the 
nodes of the ``$\epsilon$-machine" correspond to an element of a partitioning of (complete, infinite)
histories \cite{Zambella,arXiv}. An example of the latter is e.g. `rule 22' \cite{Zambella,arXiv}, 
another example is `rule 4' \cite{arXiv}.

A main point in I - III was that no such minimization can in general be expected
(and FCs are infinite), if the HMM depends on irrational real valued parameters, such that all
different observed sequences lead to different optimal forecasts. It was already pointed out in
\cite{Grass-Helv,Grass-Nat} that even rather simple systems can lead to infinite FC.
On the other hand, symmetries of the ${\bf T}^{(x)}$ can lead to non-trivial minimizations,
even if the ${\bf T}^{(x)}$ depend on irrational parameters (an instance where {\it some}
optimal forecasts do not depend on the history at all, but FC is infinite nevertheless, will be
be discussed below).
No entropy estimates with precision close to these in \cite{Zambella} were made in I,
and no estimates of FCs at all. The main reason is that no minimizations of the 
set of `mixed states' were performed.

It is claimed in I-III that the set of mixed states is in general not 
countable, if irrational parameters are involved, but this is wrong. It is always countable.
Let us discuss this very important point somewhat more in detail, because it is 
also related to the very definition of ``$\epsilon$-machines" and of causal states, and to the 
question whether the latter correspond to partitionings of history space or not --- and because it 
is also responsible for the slow progress made in "computational mechanics", whose state of 
art \footnote{Ref.~\cite{Zambella} still contains the only precise estimates of FC for non-trivial
models, \cite{Grass-Nat} gives the only such estimates for what is called ``set complexity" (SC) in
\cite{Grass86}, and \cite{Zambella} contains the only discussion of what are called ``totally recursive"
systems in \cite{arXiv}.} is in many ways still represented by \cite{Zambella,Grass-Nat}.

All these constructs were introduced \cite{Grass86,Young} with the goal of forecasting, 
more precisely with the goals of {\it performing actual} forecasts and quantifying the 
difficulties involved \footnote{According to its title, a main goal in \cite{Young} was also 
inferring complexity, but this was never seriously persued, except for \cite{Strelioff}, where  
however only extremely simple models were treated. The difficulties in inferring structural complexity are  
discussed in \cite{arXiv}.}. In such an enterprise, one cannot deal with histories extending into the 
infinite past, since they would a.s. not be finitely describable (except in trivial models). As in 
the theory of automata and formal languages \cite{Hopcroft} and as discussed above, one {\it must}
thus assume that observations started at some finite time, and one thus deals with {\it finite} 
histories. The use of infinite histories in constructing  ``$\epsilon$-machines", as done e.g. in 
\cite{Crutch_physica,Shalizi,Feldman,Loomis}, is thus based 
on a fundamental misunderstanding. From this follows directly that complete ``$\epsilon$-machines" 
must include start nodes (as, e.g. in \cite{Young}), that their nodes (the causal states) are countable, 
and that causal states correspond to equivalence classes of {\it finite} histories but to {\it coverings} 
of the set of infinite histories. The assumption in I-III and in \cite{Crutch_physica,Shalizi,Feldman,Loomis}
that the definition of causal states is based on {\it infinite} histories makes the theory more elegant
but much less useful -- and is manifestly wrong for ``totally recurrent" models \cite{Zambella,arXiv}.

Using infinite histories would lead to $\epsilon$-machines without transient parts and start nodes.
This looks like a simplification, but makes a great technical problem: Using start nodes was essential
in calculating entropies and forecasting complexities in \cite{Zambella}, `set complexities'
in \cite{Grass-Nat}, and ``transient informations” \cite{Feldman} in \cite{arXiv}. 

It is not clear when 
the transition happened from $\epsilon$-machines with start nodes and transient parts as in \cite{Young} (and 
thus with causal states corresponding to finite histories) to infinite histories and $\epsilon$-machines 
without start nodes as in \cite{Crutch_physica,Shalizi,Feldman,Loomis}. Anyhow, it was never 
mentioned.

Thus the set of ``mixed" (i.e., of correct causal) states is always countable by construction. What 
is correct, however, is that the {\it closure} of the set of mixed states (which includes states 
corresponding to infinite memory) is in many cases uncountable, just as the rational numbers are 
countable, while the reals (their closure) are not (a non-countable set could be obtained       
by using any arbitrary starting vector ${\bf P}_0$ instead of ${\bf P}_0 = {\bf P}_\infty$, but the 
resulting forecasts would then not be optimal). This non-countability of the closure is, however, not 
restricted to HMMs with real-valued parameters (as claimed in I - III), but holds also for many of the 
HMMs studied in \cite{Zambella}. The main difference between simple and complex models
is often not the cardinality of the sets of inferences, but is the Hausdorff dimension of the 
invariant measure (the Blackwell measure \cite{Blackwell}) supported by it. It was zero in all models 
studied in \cite{Zambella}, although the box-counting dimensions were non-zero in many cases.

The failure of distinguishing sets from their closures is also responsible for most of II 
being obsolete. As remarked already in \cite{Zambella}, the set of mixed (= correct causal) states is very
similar to an iterated function system (IFS), with a difference largely overlooked in I - III:
While the maps ${\bf T}^{(x)}$ are chosen randomly (maybe with 
non-uniform probabilities) in an IFS, their choice is in the present case constrained by the (in general 
non-trivial) grammar of the sequence $\{x_t\}$. Thus theorems about IFSs cannot be taken over blindly 
\footnote{Non-trivial grammatical restrictions do not seem to be responsible for the `resetting strings' 
found in some of the models studied in \cite{Zambella}, which are finite strings $R = x_1\ldots x_n$ of 
observed symbols for which ${\bf T}^{(R)} \equiv {\bf T}^{(x_n)}\ldots {\bf T}^{(x_1)}$ maps any 
${\bf P}$ onto ${\bf P}_\infty$. They correspond to catastrophic forgettings of the internal 
state and are then responsible for the finite Hausdorff dimension of the Blackwell measure; in a 
systematic search, we now found that this phenomenon occurs in 108 of the 256 models studied in 
\cite{Zambella}. For the model called `rule 22' in \cite{Zambella}. the shortest resetting string is
e.g. `01000010000' (the string given in \cite{Zambella} is an extension of this).}. Nevertheless, 
the {\it open set condition} (OSC) \cite{Moran,Peres-Rams} should hold in many cases. But its 
application is much less trivial than assumed in II, where the (non-)overlapping of the sets 
was confused with the (non-)overlapping of their closures. That such a confusion can be fatal is maybe 
best illustrated by the `fat baker's transformation' \cite{Alexander}. Another example where it leads
to a wrong conclusion, proposed originally in paper III, is discussed below. 

If minimization of the set of mixed states had been used, the study of the Hausdorff dimension of the 
set of causal states in II (called `statistical complexity dimension' there), would have 
been very useful. But although this minimization was alluded to repeatedly in I - III, 
it was never actually done.

Even more natural and interesting than the `statistical complexity dimension' seems anyhow what we 
might call the `{\it FC divergence rate}' $\alpha$, in which we replace the dependence of the FC on the
precision of the hidden state inferences (as in the statistical complexity dimension)
by its dependence on the achieved values of the Shannon entropy. Assume we have a stochastic process 
with finite entropy $h$ and infinite true FC, and assume that we have an approximate
forecasting scheme with a finite FC $C$ and an approximate entropy $h_C > h$. Then we might expect
generically
\be
    C \sim \delta h^{-\alpha}\qquad {\rm with} \;\;\; \delta h = h_C - h.
\ee
Indeed, this can be defined for general stationary ergodic processes, not only for HMM. All what is 
needed is that one can define a sequence of {\it approximate} HMMs or OOMs, to which one can apply Eqs.~(4,6).
Minimizing $h_C-h$ can be used for most accurate model (or parameter) inference \cite{Zucchini}, but minimizing
both $h_C-h$ and $C$ at fixed $\alpha$ could be useful for finding models / parameters / approximations
that are both accurate and relatively easy to forecast.

Finally, Ref. \cite{Jurgens3} (paper III) is devoted to the calculation of the {\it `ambiguity rate'} $h_a$ of a process, 
a notion supposedly introduced by Shannon in \cite{Shannon} as a quantity opposed to `equivocation'. In fact, 
`ambiguity' is mentioned exactly once in \cite{Shannon}, but as a {\it synonym} for equivocation. The 
concept as defined in III is the rate at which information about the past can be 
discarded, and yet prediction of the future is optimal. If $h_a< h$, this would mean that less information
is dismissed than new information is obtained, i.e. that FC (the information stored in an optimal forecasting 
scheme) diverges with time. In the opposite case, i.e. when $h_a= h$, it was claimed in III that ``the associated
$\epsilon$-machine's internal causal-state process is stationary." This is wrong, as 
demonstrated by the counter-examples given in \cite{Zambella}. There it was shown that FC is finite in 
all considered cases (i.e., $h_a= h$), but yet the forecasting process is not stationary in cases like 
`rule 22', because of the slow convergence of FC. 

When each observed sequence leads to a different forecast, it is clear that no information can be 
dismissed in an optimal forecast, and thus $h_a = 0$. In some cases, the situation is a bit more subtle.
Consider, e.g., the example in Sec. 7B of
III for control parameter values $x=0.25, \alpha= 0.5$, which was treated in detail in 
appendix B of III. In this model, there are three emitted symbols $\triangle,\Box,\circ$. 
One finds that always $\hat{p}_t(\Box)=1/4$, while $\hat{p}_t(\triangle)$ and $\hat{p}_t(\circ)$ 
depend non-trivially on the observations at times $1,\ldots t-1$. Indeed, no two different histories lead
to the same $\hat{p}_t(\triangle)$ and $\hat{p}_t(\circ)$ (at least for $t<12$; for larger $t$ one would
need extended precision arithmetic to avoid numerical problems). For each $t$, the largest values of 
$\hat{p}_t(\triangle)$ and $\hat{p}_t(\circ)$ coincide, and so do the smallest values. The difference 
between the two increases with $t$ and tends to $0.07020(1)$ for $t\to\infty$, which shows clearly that 
the history dependence
does not vanish asymptotically. Thus, no information can be dismissed if {\it all} forecasts 
are to be optimal, and $h_a = 0$ --- while  $h_a = 0.4499$ bits is quoted in III. 
Obviously, this wrong value was obtained because $h_a$ was calculated in III using 
the open set condition, and overlaps between sets were confused with overlaps between their closures. 
Since this error was made in all calculations of $h_a$, all numerical results of 
III are either trivial or obsolete.

Acknowledgements: I thank Sarah Marzen and Paul Riechers for extremely useful discussions.

\end{document}